# Classical and quantum dynamics of the $n$-dimensional kicked rotator

Georg Junker, Harald Karl*, and Hajo Leschke
*Institut für Theoretische Physik, Universität Erlangen-Nürnberg,*
*Staudtstr. 7, D-91058 Erlangen, Germany*

The classical and quantum dynamics for an $n$-dimensional generalization of the kicked planar ($n = 1$) rotator is considered. The classical behavior of this system is, in essence, described by a one-dimensional kicked rotator in an additional effective centrifugal potential. Therefore, typical phenomena like the diffusion in classical phase space are similar to that of the one-dimensional model. For the quantum dynamics such a result is not expected as in this case the evolution does depend in a very complicated way on the number $n$ of degrees of freedom. In the limit $n \to \infty$ we find the free undisturbed quantum motion. For finite values of $n$ ($1 \leq n \leq 26$) we study numerically the quantum dynamics. Here, we always find localization independent of the actual number of degrees of freedom.

## I. INTRODUCTION

Quantum signatures of classically chaotic systems have attracted increasing attention in the last 10-15 years [1–5]. One particular class of systems which has been investigated is characterized by time-periodic Hamiltonians [6]. The most prominent member of this class is the so-called kicked planar rotator. This is a point particle which moves freely along the unit circle in the plane. This motion is perturbed by time-periodic kicks whose strength does depend on the actual position of the particle. Accordingly, the Lagrangian of this system may be written as

$$L_1(\vartheta, \dot{\vartheta}, t) = \frac{I}{2}\dot{\vartheta}^2 - K\cos\vartheta \sum_{j=-\infty}^{+\infty} \delta\left(\frac{t}{\tau} - j\right). \qquad (1)$$

Here the angle $\vartheta \in [0, 2\pi)$ describes the position of the particle and $I$ stands for its momentum of inertia with respect to the origin. Furthermore, $\tau > 0$ is the period of the kicks and $K > 0$ is the maximal kicking strength. It is this "stochasticity" parameter $K$ which controls the classical dynamical behavior of the system. Near $K\tau^2/I \approx 0.9716\ldots$ there is a transition from local to global stochasticity taking place [7]. This is typically visible in the linear increase of the system's energy in time.

In the present paper we will introduce an $n$-dimensional generalization of the system characterized by (1) and study its classical as well as its quantum dynamics. More precisely, we consider the free motion of a point mass on the $n$-dimensional unit sphere embedded in the $(n+1)$-dimensional Euclidean space. This motion is periodically disturbed by kicks, whose strength depends only on one of the $n+1$ Cartesian coordinates. On the classical level this generalized system is essentially identical to the one-dimensional case, as the effective equation of motion does not depend on the number of degrees of freedom. This is in contrast to the quantum dynamics of the $n$-dimensional kicked rotator, which explicitly depends in a rather complicated way on the number $n$. The study of this $n$-dependence has been our primary motivation for this work. Let us also mention that our model for $n = 2$ is somehow related to the kicked top extensively discussed by Haake and coworkers [8]. See also ref. [5].

Our paper is organized as follows. In the next Section we explicitly define the $n$-dimensional generalization of (1) which we are going to investigate. Section 3 is devoted to the study of the classical dynamics. We integrate the classical equation of motion and obtain a certain $n$-dependent generalization of the well-known standard map. The observed diffusive behavior of this map is compared with that for the one-dimensional case. In Section 4 we will present analytical as well as numerical results for the quantum dynamics. In the limit $n \to \infty$ the $n$-dimensional kicked rotator behaves like a free rotator (without kicks). For finite values of $n$ we always observe — by numerical methods — the phenomenon of so-called localization. The corresponding localization length is, in contrast to what one might expect, found to be independent of $n$ up to values as large as 26. Unfortunately, our numerical method does not allow to draw firm conclusions for even larger values of $n$.

*Present address: Fachbereich Physik, Universität - Gesamthochschule - Essen, Universitätsstr. 5, D-45117 Essen, Germany

## II. THE MODEL

The $n$-dimensional generalization of the planar rotator is simply the free motion of a point mass on the unit sphere embedded in the $(n+1)$-dimensional Euclidean space

$$S^n := \left\{ (x_1, \dots, x_{n+1}) \in \mathbb{R}^{n+1} \,\middle|\, \sum_{r=1}^{n+1} x_r^2 = 1 \right\}. \qquad (2)$$

The unit sphere $S^n$ is usually parametrized by polar coordinates:

$$\begin{aligned} x_r &= e_r \sin\vartheta, \qquad r = 1, 2, \dots, n, \\ z \equiv x_{n+1} &= \cos\vartheta, \qquad 0 \le \vartheta \le \pi. \end{aligned} \qquad (3)$$

Here $\vartheta$ denotes the angle between the unit vector $\vec{x} := (x_1, \dots, x_{n+1})$ and the $z$-axis which is chosen to be the coordinate axis for the last component of $\vec{x}$. In the above $e_i$ stands for the $i$-th component of an arbitrary unit vector $\vec{e} := (e_1, \dots, e_n, 0) \in \mathbb{R}^{n+1}$ perpendicular to the $z$-axis. Obviously, $\dot{\vec{x}}^2 = \dot\vartheta^2 + \dot{\vec{e}}^2 \sin^2\vartheta$ is the squared modulus of the velocity of a point particle moving along a path $\vec{x} = \vec{x}(t) \in S^n$.

The Lagrangian of the $n$-dimensional kicked rotator is then defined by

$$\begin{aligned} L_n(\vartheta, \dot\vartheta, \dot{\vec{e}}, t) = \; & \frac{I}{2}\left(\dot\vartheta^2 + \dot{\vec{e}}^2 \sin^2\vartheta\right) \\ & - K \cos\vartheta \sum_{j=-\infty}^{\infty} \delta\left(\frac{t}{\tau} - j\right). \end{aligned} \qquad (4)$$

As in (1), $I$ is the moment of inertia, $K$ is the kicking strength and $\tau$ the period of these kicks. Due to the rotational invariance of (4) about the $z$-axis we have $n-1$ constants of motion:

$$\vec{M} := \partial L_n / \partial \dot{\vec{e}} = I \dot{\vec{e}} \sin^2\vartheta \,. \qquad (5)$$

They allow the reduction of system (4) to an effective one-dimensional system, which in turn allows for a solution of the classical equations of motion in terms of a time-periodic classical map in a reduced phase space. We will derive this $n$-dimensional generalization of the standard map in the next Section.

## III. THE CLASSICAL DYNAMICS

Due to the constants of motion (5) the dynamics of the system (4) can be reduced to a one-dimensional system characterized by the effective Lagrangian

$$\begin{aligned} L_{eff}(\vartheta, \dot\vartheta, t) = \; & \frac{I}{2}\dot\vartheta^2 - \frac{M^2}{2I \sin^2\vartheta} \\ & - K \cos\vartheta \sum_{j=-\infty}^{\infty} \delta\left(\frac{t}{\tau} - j\right), \end{aligned} \qquad (6)$$

where $M := |\vec{M}|$. Let us note that for the special case $M = 0$ the above Lagrangian is identical to that for the planar rotator. Indeed, classically the case $M = 0$ cannot be distinguished from the system (1) as in the former case the motion takes place in a plane containing the $z$-axis. Let us add, that in this case the parametrization (3) is singular and should be replaced by $0 \le \vartheta < 2\pi$. In the general case $M > 0$ the Lagrangian (6) may be interpreted as that for the kicked planar rotator in an additional time-independent (effective) potential.

The classical equation of motion, which may be derived form $L_{eff}$, can easily be reduced to a discrete-time map on the two-dimensional classical phase space $\Omega := [0, \pi] \times \mathbb{R}$. This is because (6) describes a regular motion between two successive kicks. Introducing the quantities $z_j := z(j\tau) = \cos\vartheta(j\tau)$, the position ($z$-coordinate) at the $j$-th kick, and $\dot z_j := \lim_{\varepsilon\downarrow 0} \dot z(j\tau + \varepsilon)$, the corresponding velocity just after the $j$-th kick, the equation of motion leads to the phase-space map ($\vartheta_j := \arccos z_j$, $\dot\vartheta_j := -\dot z_j / \sin\vartheta_j$)

$$T : \left\{ \begin{array}{l} \Omega \to \Omega \\ (\vartheta_j, \dot\vartheta_j) \mapsto (\vartheta_{j+1}, \dot\vartheta_{j+1}) \end{array} \right. . \qquad (7)$$

This map is explicitly given by

$$\begin{aligned} z_{j+1} &= A_j \sin\left[\omega_j \tau \,\mathrm{sgn}(\dot z_j) + \phi_j\right], \\ \dot z_{j+1} &= \omega_j A_j \,\mathrm{sgn}(\dot z_j) \cos\left[\omega_j \tau \,\mathrm{sgn}(\dot z_j) + \phi_j\right] \\ & \quad - \kappa(1 - z_{j+1}^2), \end{aligned} \qquad (8)$$

where we have set

$$\begin{aligned} A_j &:= \sqrt{1 - \frac{M^2}{I^2 \omega_j^2}}, \qquad \omega_j := \sqrt{\frac{(M/I)^2 + (\dot z_j)^2}{1 - z_j^2}}, \\ \phi_j &:= \arcsin\left(\frac{z_j}{A_j}\right), \qquad \mathrm{sgn}(z) := \left\{ \begin{array}{ll} z/|z| & \text{for } z \neq 0 \\ 1 & \text{for } z = 0 \end{array} \right. \end{aligned} \qquad (9)$$

and introduced the rescaled kicking strength $\kappa := K\tau/I$. Note that

$$\lim_{\varepsilon\downarrow 0}\Big(\dot z(j\tau + \varepsilon) - \dot z(j\tau - \varepsilon)\Big) = \kappa\left(z_j^2 - 1\right). \qquad (10)$$

We leave it to the reader to show that for $M = 0$ the map (7), respectively, (8) reduces to the well-known standard map for the planar rotator.

We have iterated the map (7) numerically for various values of the parameters $\kappa$ and $M$ using units such that $I = \tau = 1$. Figures 1-3 show typical phase-space portraits generated by the above map. It is obvious that with an increasing value of $\kappa$, keeping $M$ fixed, the islands of regular motion become smaller. Whereas, an increasing $M$ for fixed $\kappa$ will cause growing islands of stability. In particular, Figure 1 displays, because of $M = 0$, the well-known phase-space portrait of the kicked planar rotator.

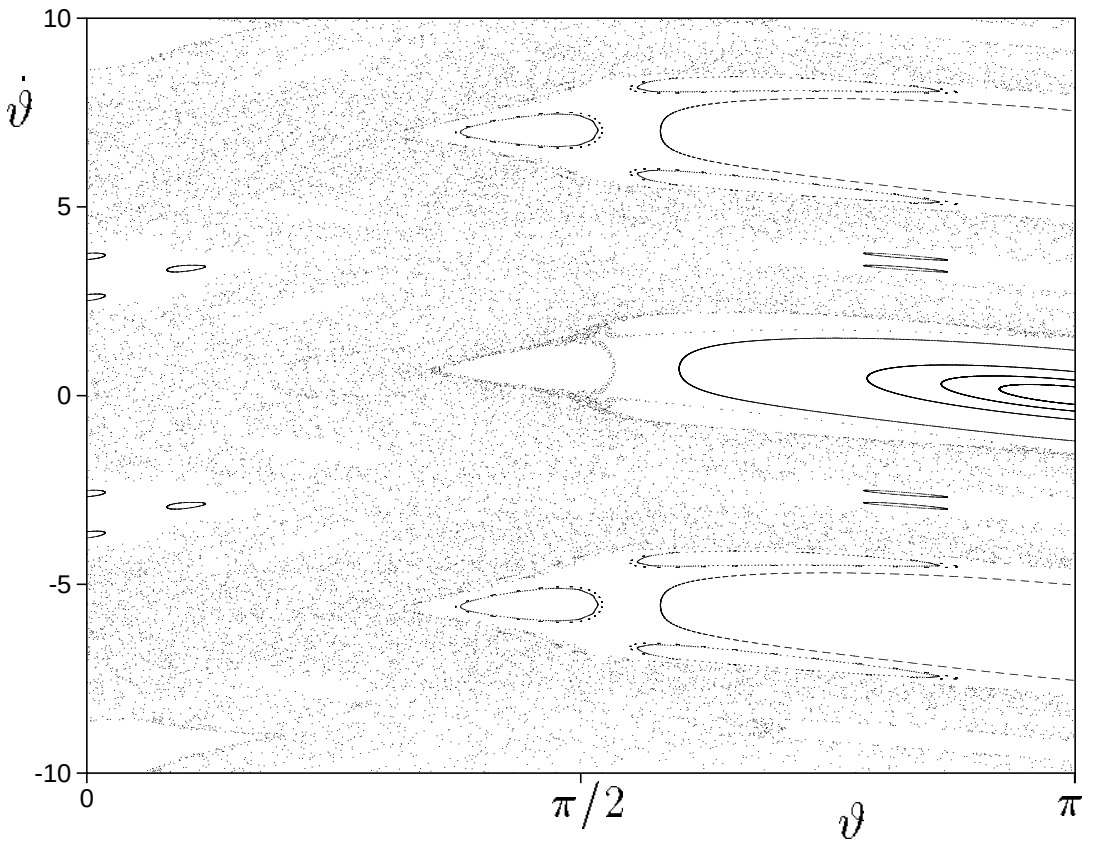

FIG. 1. Phase-space portrait generated by the map (7) for parameter $\kappa = 1.5$ and $M = 0$.

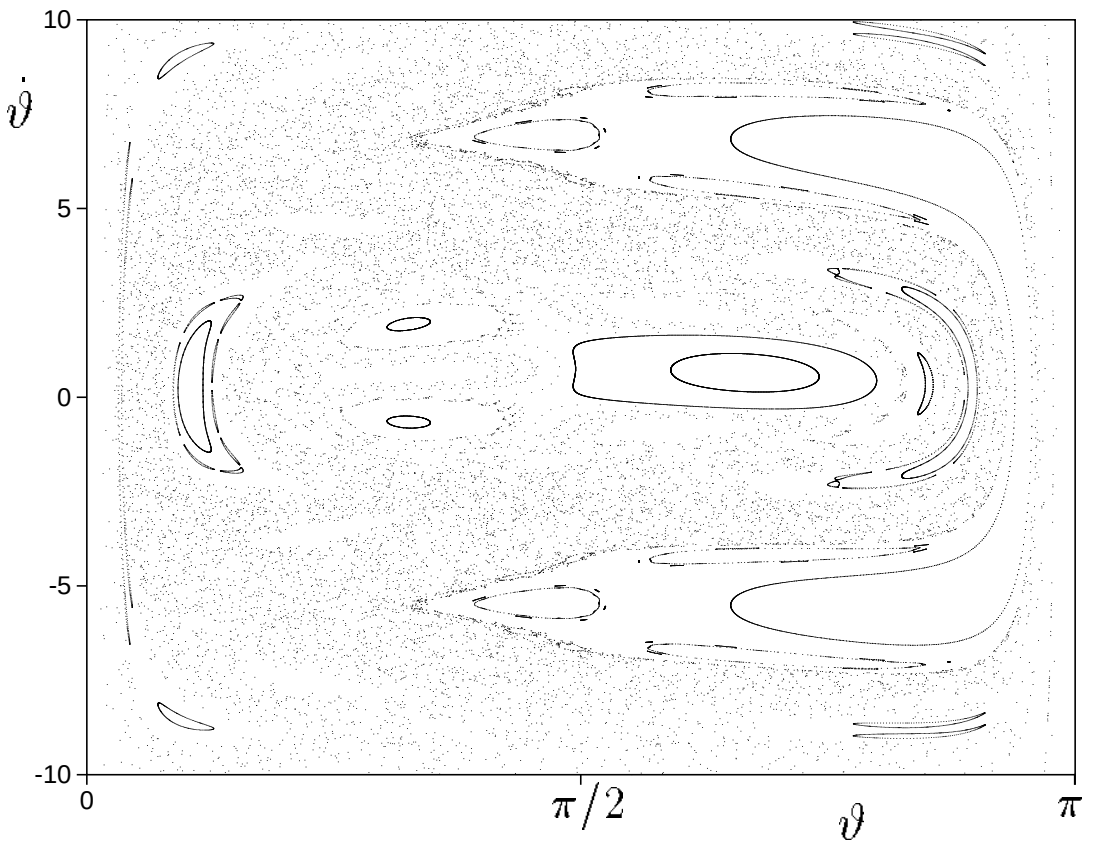

FIG. 2. Same as Figure 1 for $\kappa = 1.5$ and $M = 1$.

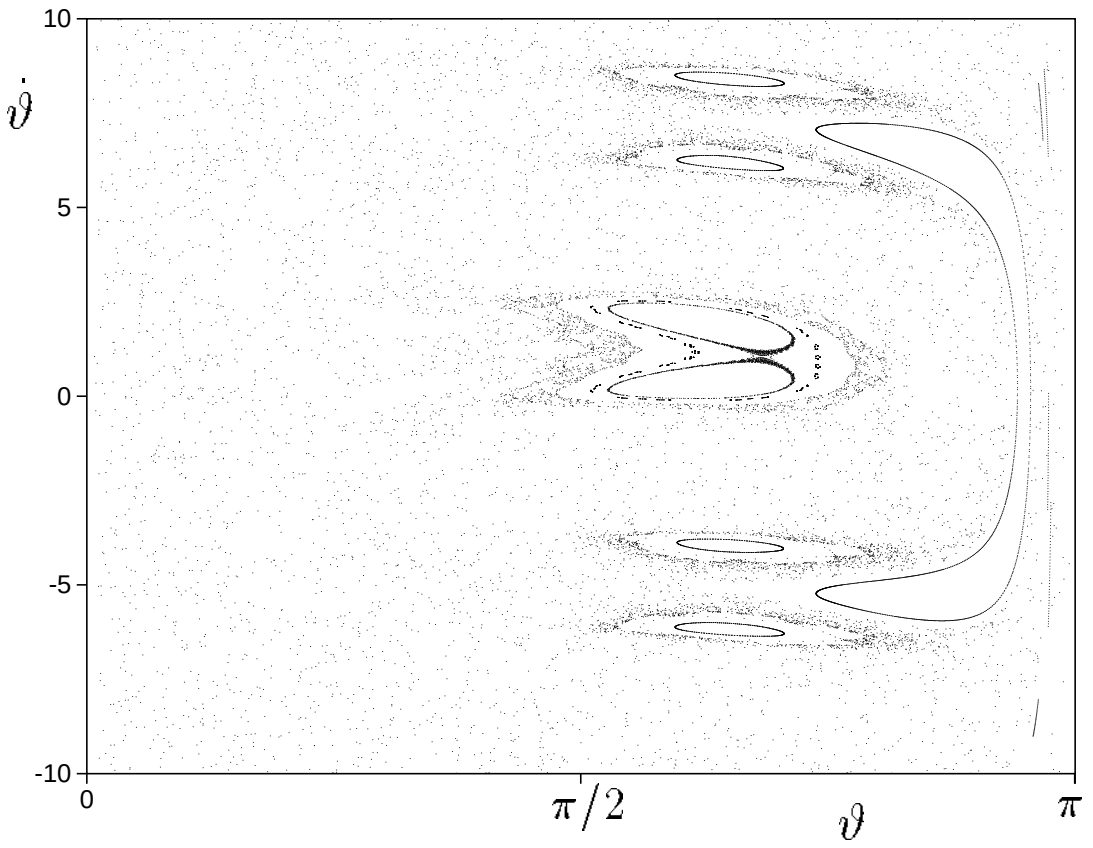

FIG. 3. Same as Figure 1 for $\kappa = 2.5$ and $M = 1$.

In order to study the stochastic nature of the map (7) quantitatively we have looked at the diffusive growth of the energy $E_j := I\omega_j^2/2$. For sufficiently large $\kappa$ it is expected that a random-phase approximation is valid [7]. That is, one assumes a uniform distribution for $\vartheta_j$ in the interval $[0, \pi]$ and an independent distribution for $\dot\vartheta_j$ which is symmetric about the origin. These assumptions directly lead to a linear (hence diffusive) growth of the energy with time. In other words, the diffusion coefficient

$$D(\kappa) := \lim_{j\to\infty} (E_j - E_0)/j \tag{11}$$

exists. Its leading dependence on $\kappa$ for large $\kappa$ is given by

$$D(\kappa) \sim I\kappa^2/4 =: D_0(\kappa), \qquad \kappa \to \infty. \tag{12}$$

For the calculation of subleading corrections one has to go beyond the random-phase approximation. For the special case $M = 0$ this has been done by Rechester and White [9] (cf. also [7]) with the result

$$D(\kappa)/D_0(\kappa) \sim 1 - 2\Big(\mathrm{J}_2(\kappa) + \mathrm{J}_1^2(\kappa) - \mathrm{J}_2^2(\kappa) - \mathrm{J}_3^2(\kappa)\Big), \tag{13}$$

where $\mathrm{J}_\nu$ denotes the Bessel function with index $\nu$.

We have calculated the $\kappa$ dependence of the diffusion coefficient (11) for various values of $M$ with the result that $D(\kappa)$ is well described by (13) independent of the parameter $M$. Figures 4 and 5 present our results for $M = 0$ and $M = 10$, respectively. The observed enhancement in the diffusion coefficient near $\kappa \approx 2\pi m$, $m \in \mathbb{N}$, is attributed to so-called acceleration modes [10].

In concluding this Section we may say that the classical $n$-dimensional kicked rotator is, as far as its stochastic nature is concerned, identical to the one-dimensional kicked rotator.

## IV. THE QUANTUM DYNAMICS

Let us now consider the quantum version of the model introduced in Section II. It is characterized by the Hamiltonian:

$$H_n := -\frac{\hbar^2}{2I}\,\Delta + K\cos\Theta \sum_{j=-\infty}^{\infty} \delta\left(\frac{t}{\tau} - j\right). \tag{14}$$

Here $\Delta$ stands for the Laplace-Beltrami operator defined on the unit sphere $S^n$ and $\Theta$ represents the position operator corresponding to the polar angle $\vartheta$.

Before we study the unitary time evolution generated by (14) we comment on the operator $\Delta$ and the Hilbert space $L^2(S^n)$ it is acting on. It is well known [11] that $L^2(S^n)$ can uniquely be decomposed into invariant orthogonal subspaces $\mathcal{D}^\ell$ each of which carries a unitary irreducible representation of the group $SO(n+1)$ labeled, in analogy to $SO(3)$, by $\ell = 0, 1, 2, \ldots$:

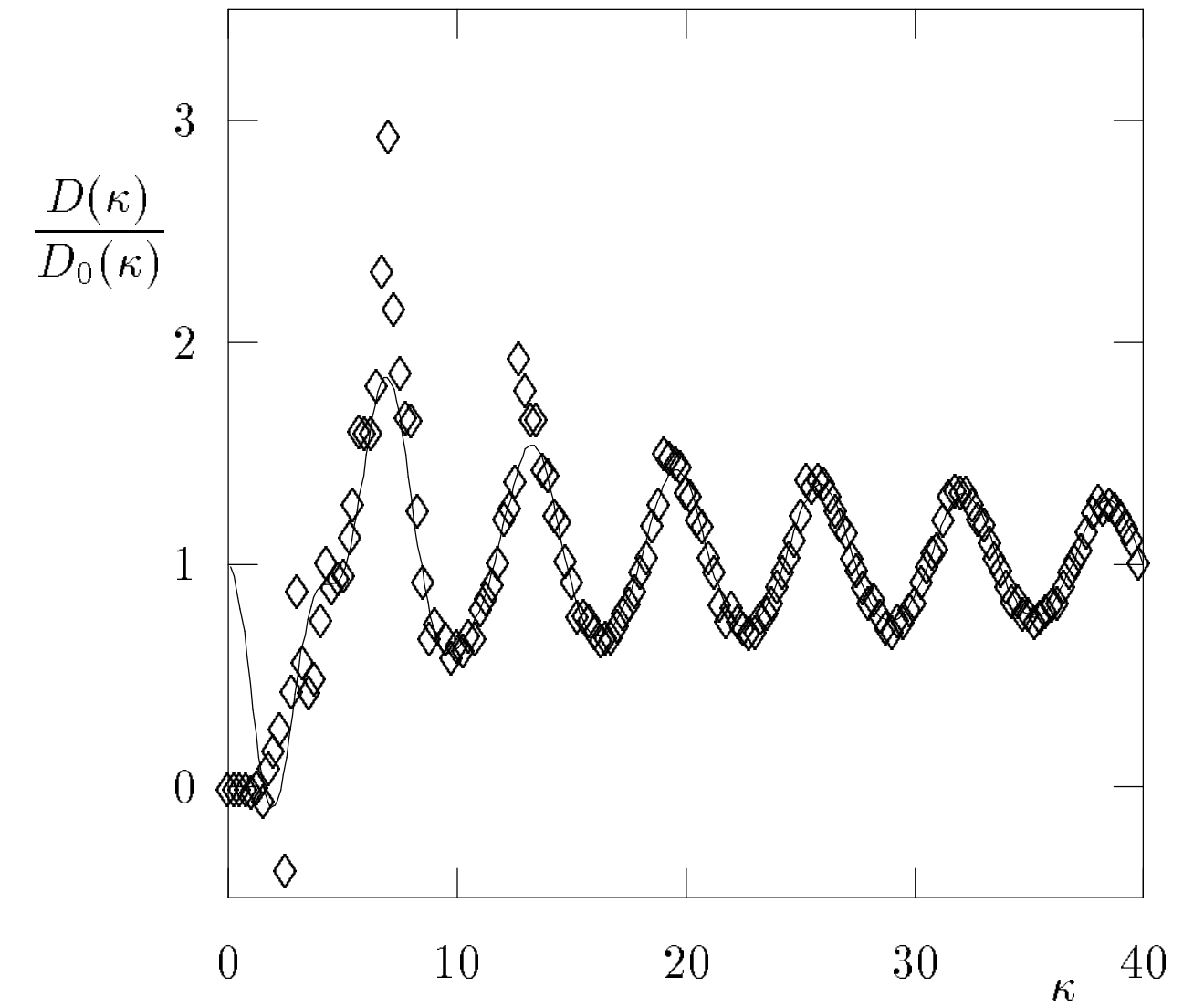

FIG. 4. The diffusion coefficient $D(\kappa)/D_0(\kappa)$ for $M = 0$. The solid line shows the expected analytical behavior (13).

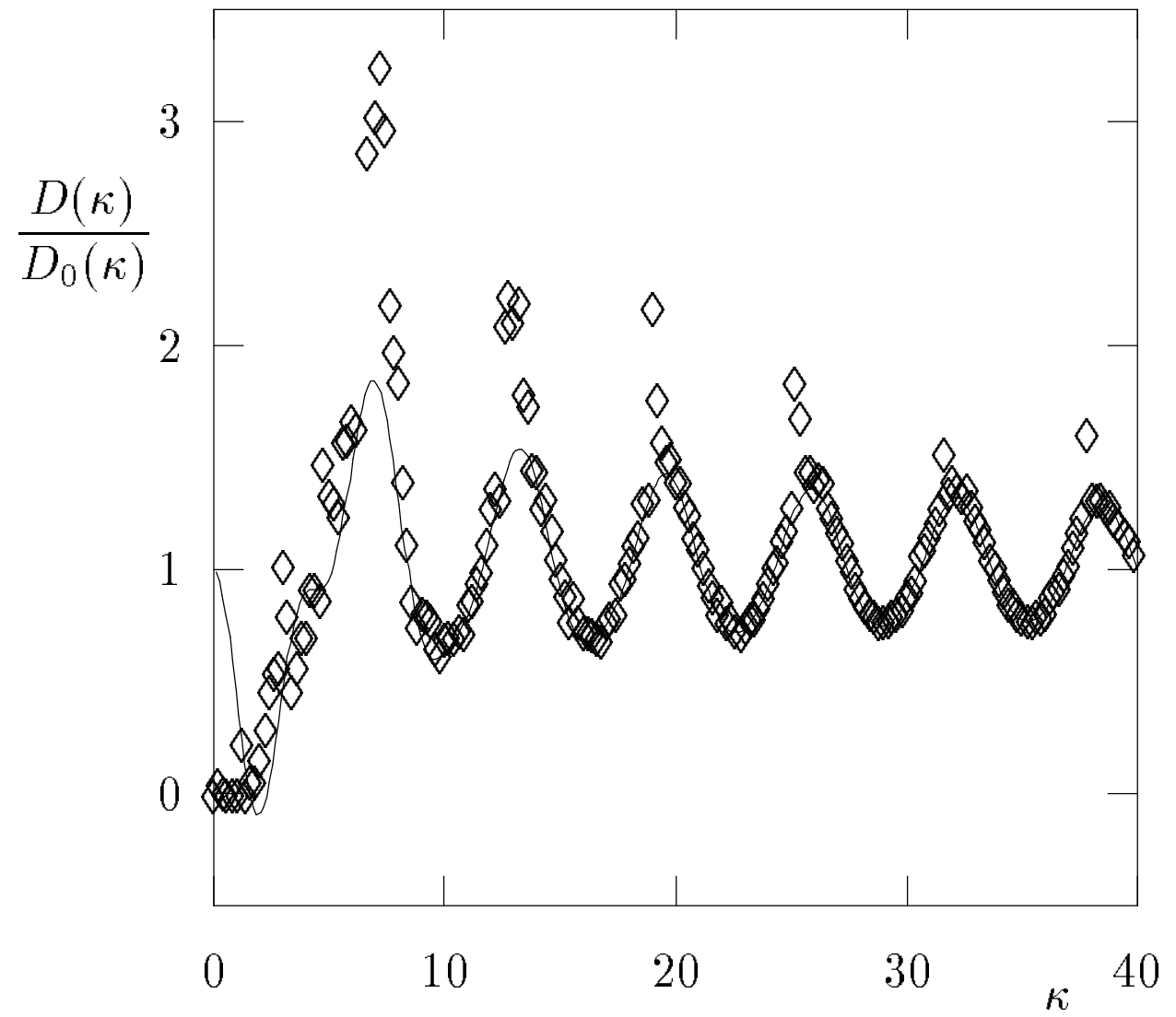

FIG. 5. Same as Figure 4 for parameter $M = 10$.

$$L^2(S^n) = \bigoplus_{\ell=0}^{\infty} \mathcal{D}^\ell. \tag{15}$$

The $\ell$-th subspace has the finite dimension

$$d_\ell := (2\ell + n - 1)\,\frac{\Gamma(\ell + n - 1)}{\Gamma(\ell + 1)\Gamma(n)}, \tag{16}$$

where $\Gamma$ denotes Euler's gamma function. In fact, the subspace $\mathcal{D}^\ell$ is the eigenspace of $\Delta$ corresponding to the eigenvalue $-\ell(\ell + n - 1)$, see ref. [11]. An orthonormal basis in $\mathcal{D}^\ell$ is the common eigenbasis $\{|\ell\mathcal{M}\rangle\}_{\mathcal{M}}$ of the $n - 1$ commuting Casimir operators of the subgroups $SO(p)$, $p = 2, \ldots, n$, appearing in the group chain $SO(2) \subset \cdots \subset SO(n) \subset SO(n + 1)$ [12]. Here $\mathcal{M}$ stands for the $(n - 1)$-tuple

$$\mathcal{M} := (m_1, m_2, \cdots, m_{n-2}, m_{n-1}) \tag{17}$$

where

$$|m_1| \le m_2 \le \cdots \le m_{n-2} \le m_{n-1} \le \ell \tag{18}$$

with $m_1 \in \mathbb{Z}$ and $m_r \in \mathbb{N}_0$ for $r = 2, \cdots, n - 1$. The spectral decomposition of $\Delta$ now reads

$$\Delta = -\sum_{\ell=0}^{\infty} \ell(\ell + n - 1) \sum_{\mathcal{M}} |\ell\mathcal{M}\rangle\langle\ell\mathcal{M}|. \tag{19}$$

With the above remarks we can discuss the time evolution generated by the Hamiltonian (14). Let us first consider the one-kick evolution operator

$$U := \exp\left\{\frac{i\hbar\tau}{2I}\,\Delta\right\} \exp\left\{-\frac{i}{\hbar}\,K\tau\cos\Theta\right\}, \tag{20}$$

which describes the time evolution for one period $\tau$ starting right before a kick. If $|\Psi_j\rangle$ denotes a state right before the $j$-th kick the time evolution of this state is given by

$$|\Psi_{j+1}\rangle = U|\Psi_j\rangle. \tag{21}$$

The properties of this quantum map are our main concern.

In order to make (21) more explicit we expand the state $|\Psi_j\rangle$, for each $j$, into the basis $\{|\ell\mathcal{M}\rangle\}$:

$$|\Psi_j\rangle = \sum_{\ell,\mathcal{M}} a_{\ell\mathcal{M}}(j)\,|\ell\mathcal{M}\rangle \tag{22}$$

where

$$a_{\ell\mathcal{M}}(j) := \langle\ell\mathcal{M}|\Psi_j\rangle. \tag{23}$$

Rewriting (21) in terms of these expansion coefficients yields the recurrence relation

$$\begin{aligned} a_{\ell\mathcal{M}}(j+1) &= \exp\left\{-\frac{i\hbar\tau}{2I}\,\ell(\ell+n-1)\right\} \\ &\times \sum_{\ell',\mathcal{M}'} a_{\ell'\mathcal{M}'}(j)\,\langle\ell\mathcal{M}|\exp\{-(i/\hbar)K\tau\cos\Theta\}\,|\ell'\mathcal{M}'\rangle. \end{aligned} \tag{24}$$

What remains to be done is the explicit calculation of the matrix element appearing in (24). In doing so we will work in the coordinate representation

$$Y_{\ell\mathcal{M}}(\vartheta, \vec{e}) := \langle\vartheta, \vec{e}\,|\ell\mathcal{M}\rangle, \tag{25}$$

that is, with the $n$-dimensional generalization of the spherical harmonics. They are orthonormalized according to

$$\int_{S^n} d\vartheta \sin^{n-1}\vartheta \, d\vec{e}\, Y^*_{\ell'\mathcal{M}'}(\vartheta,\vec{e}) Y_{\ell\mathcal{M}}(\vartheta,\vec{e}) = \delta_{\ell,\ell'}\delta_{\mathcal{M},\mathcal{M}'}. \tag{26}$$

Here and below we will adopt the notation and results of ref. [13]. In terms of the spherical harmonics (25) the matrix element reads

$$\begin{aligned} &\langle \ell\mathcal{M}|\exp\{-(i/\hbar)K\tau\cos\Theta\}|\ell'\mathcal{M}'\rangle \\ &= \int_{S^n} d\vartheta \sin^{n-1}\vartheta\, d\vec{e}\, \exp\{-(i/\hbar)K\tau\cos\vartheta\} \\ &\times Y^*_{\ell\mathcal{M}}(\vartheta,\vec{e}) Y_{\ell'\mathcal{M}'}(\vartheta,\vec{e}). \end{aligned} \tag{27}$$

In order to make this integral somewhat more explicit we expand the exponential function in terms of spherical harmonics:

$$\exp\{-(i/\hbar)K\tau\cos\vartheta\} = \sum_{\ell''=0}^{\infty} Y_{\ell''\mathcal{O}}(\vartheta,\vec{e})\, B_{\ell''}(\lambda), \tag{28}$$

where we have introduced $\lambda := K\tau/\hbar$ for brevity. Because of its invariance under rotations about the $z$-axis only those spherical harmonics with $\mathcal{M}'' = \mathcal{O} := (0,\ldots,0)$ contribute to the expansion of this exponential. The expansion coefficient is explicitly given by

$$\begin{aligned} B_\ell(\lambda) &:= \int_{S^n} d\vartheta \sin^{n-1}\vartheta\, d\vec{e}\, Y^*_{\ell\mathcal{O}}(\vartheta,\vec{e}) \exp\{-i\lambda\cos\vartheta\} \\ &= \sqrt{d_\ell |S^n|}\,\Gamma\left(\frac{n+1}{2}\right)(-i)^{\ell/2}(2/\lambda)^{(n-1)/2} \\ &\quad\times \mathrm{J}_{\ell+(n-1)/2}(\lambda). \end{aligned} \tag{29}$$

For the integration we have made use of the fact that $Y_{\ell\mathcal{O}}(\vartheta,\vec{e})$ is expressible in terms of a Gegenbauer polynomial. See, for example, ref. [11].

In a second step we make use of the equality

$$\begin{aligned} Y_{\ell'\mathcal{M}'}(\vartheta,\vec{e}) Y_{\ell''\mathcal{O}}(\vartheta,\vec{e}) &= \sqrt{\frac{d_{\ell'}d_{\ell''}}{|S^n| d_\ell}} \sum_{\ell\mathcal{M}} Y_{\ell\mathcal{M}}(\vartheta,\vec{e}) \\ &\times \langle \ell'\mathcal{O};\ell''\mathcal{O}|(\ell'\ell'')\ell\mathcal{O}\rangle \langle (\ell'\ell'')\ell\mathcal{M}|\ell'\mathcal{M}';\ell''\mathcal{O}\rangle \end{aligned} \tag{30}$$

which can be derived from eq. (34) in connection with eq. (15) of ref. [13]. Here $\langle \ell_1\mathcal{M}_1;\ell_2\mathcal{M}_2|(\ell_1\ell_2)\ell\mathcal{M}\rangle$ denotes a generalized Clebsch-Gordan coefficient and $|S^n| := 2\pi^{(n+1)/2}/\Gamma(\frac{n+1}{2})$ stands for the volume of the unit sphere $S^n$. Making use of the orthogonality relation (26) we finally arrive at the result

$$\begin{aligned} \langle \ell\mathcal{M}|\exp\{-(i/\hbar)K\tau\cos\Theta\}|\ell'\mathcal{M}'\rangle &= \sum_{\ell''=0}^{\infty}\sqrt{\frac{d_{\ell'}d_{\ell''}}{|S^n| d_\ell}} \\ \times B_{\ell''}(\lambda)\langle \ell'\mathcal{O};\ell''\mathcal{O}|(\ell'\ell'')\ell\mathcal{O}\rangle &\langle (\ell'\ell'')\ell\mathcal{M}|\ell'\mathcal{M}';\ell''\mathcal{O}\rangle. \end{aligned} \tag{31}$$

Using the fact that the Clebsch-Gordan coefficient $\langle \ell'\mathcal{M}';\ell''\mathcal{O}|(\ell'\ell'')\ell\mathcal{M}\rangle$ vanishes unless $\mathcal{M} = \mathcal{M}'$ we can put the map (24) into the simple form

$$\begin{aligned} a_{\ell\mathcal{M}}(j+1) &= \exp\left\{-\frac{i\hbar\tau}{2I}\,\ell(\ell+n-1)\right\} \\ &\times \sum_{\ell',\ell''=0}^{\infty} a_{\ell'\mathcal{M}}(j)\, B_{\ell''}(\lambda) G_{\mathcal{M}}(\ell',\ell'';\ell), \end{aligned} \tag{32}$$

where we have introduced

$$\begin{aligned} G_{\mathcal{M}}(\ell',\ell'';\ell) := \sqrt{\frac{d_{\ell'}d_{\ell''}}{|S^n| d_\ell}}\,&\langle \ell'\mathcal{O};\ell''\mathcal{O}|(\ell'\ell'')\ell\mathcal{O}\rangle \\ &\times\langle (\ell'\ell'')\ell\mathcal{M}|\ell'\mathcal{M};\ell''\mathcal{O}\rangle. \end{aligned} \tag{33}$$

This quantity is of purely geometric origin and, in essence, controls which angular-momentum eigenstates $|\ell\mathcal{M}\rangle$ can be reached from an initial eigenstate $|\ell'\mathcal{M}\rangle$ due to an angular-momentum transfer $\ell''$ stemming from the kick. In contrast to this, the term $B_{\ell''}(\lambda)$ is of dynamical origin and represents the weight with which the kick contributes to the angular-momentum transfer.

The map (32) also explicates that the $(n-1)$-tuple $\mathcal{M}$ is conserved. That means, there exists $n-1$ constants of motion. These are in fact the counterparts of the $n-1$ classical constants of motion (5). In particular, selecting an initial state $|\Psi_0\rangle$ such that $a_{\ell\mathcal{M}}(0) \propto \delta_{\mathcal{M},\mathcal{O}}$ the map (32) represents the quantum version of the classical map (8) with $M = 0$, hence, the well-known standard map. Whereas (8) does not depend on the number $n$ of degrees of freedom, the quantum map (32) explicitly depends in a rather complicated way on $n$. With (32) we have $\mathbb{N}$ different quantum versions for the classical standard map.

Let us note that in principal we are now in a position where we could study the map (32) numerically. However, the explicit calculation for the Clebsch-Gordan coefficients is rather complicated and also requires an extensive numerical effort [13]. Only the special initial condition $\mathcal{M} = \mathcal{O}$ allows for a practicable numerical iteration as in this case the Clebsch-Gordan coefficients are available in closed form [13]. Hence, in the following we will concentrate on this special case.

### A. The special case $\mathcal{M} = \mathcal{O}$ and the large $n$ limit

In this Section we consider the quantum map (32) with the special initial condition $\mathcal{M} = \mathcal{O}$. For convenience we use the notation $a_\ell(j) := a_{\ell\mathcal{O}}(j)$. As we have already mentioned, the Clebsch-Gordan coefficients for this case are known in closed form [13]. Hence, we are also able to express $G_{\mathcal{O}}(\ell_1,\ell_2;\ell_3)$ in a closed form:

$$\begin{aligned} G_{\mathcal{O}}(\ell_1,\ell_2;\ell_3) = &\frac{\Gamma(J+n-1)}{\sqrt{|S^n|}\,\Gamma^2\left(\frac{n+1}{2}\right)\Gamma(n-1)\Gamma\left(J+\frac{n+1}{2}\right)} \\ &\times\prod_{i=1}^{3}\left[\frac{\left(\ell_i+\frac{n-1}{2}\right)}{\sqrt{d_{\ell_i}}}\,\frac{\Gamma\left(J-\ell_i+\frac{n-1}{2}\right)}{\Gamma(J-\ell_i+1)}\right]. \end{aligned} \tag{34}$$

This expression is only valid if $J := (\ell_1+\ell_2+\ell_3)/2 \in \mathbb{N}_0$ and the $\ell_i$ obey the triangular relation $\ell_j + \ell_k \geq \ell_i \geq |\ell_j - \ell_k|$, $i, j, k \in \{1, 2, 3\}$. This conditions are similar to those known for the case $n = 2$. Whenever $J$ is not an integer or the triangular relation is not fulfilled the Clebsch-Gordan coefficients and hence $G_{\mathcal{O}}(\ell_1, \ell_2; \ell_3)$ vanishes.

With this result the map (32) reduces to

$$\begin{aligned} a_\ell(j+1) = &\ \exp\left\{-\frac{i\hbar\tau}{2I}\,\ell(\ell+n-1)\right\} \\ &\times \sum_{\ell',\ell''=0}^{\infty} a_{\ell'}(j)\, B_{\ell''}(\lambda)\, G_{\mathcal{O}}(\ell',\ell'';\ell). \end{aligned} \tag{35}$$

This recurrence relation together with the explicit expression (34) is now in a form which allows for a numerical iteration.

However, before we are going to report our numerical results let us briefly consider the limit $n \to \infty$. Using the asymptotic form of the Bessel function for large index and the Stirling formula applicable for large arguments of the gamma function one finds the following leading-order terms for $n \to \infty$:

$$\begin{aligned} B_{\ell''}(\lambda) &\approx \sqrt{|S^n|d_{\ell''}}\,(-i\lambda/n)^{\ell''}, \\ G_{\mathcal{O}}(\ell',\ell'';\ell) &\approx \frac{n^{\ell''/2}}{\sqrt{|S^n|d_{\ell''}}}\frac{\sqrt{\Gamma(\ell'+1)\Gamma(\ell+1)}}{\Gamma(J-\ell+1)} \\ &\quad\times\frac{1}{\Gamma(J-\ell'+1)\Gamma(J-\ell''+1)}. \end{aligned} \tag{36}$$

Inserting these approximations into (35) results in

$$\begin{aligned} &a_\ell(j+1) \approx \exp\left\{-\frac{i\hbar\tau}{2I}\,\ell(\ell+n-1)\right\} \\ &\times \sum_{\ell',\ell''=0}^{\infty} a_{\ell'}(j)\frac{(-i)^{\ell''/2}(\lambda/\sqrt{n})^{\ell''}\sqrt{\Gamma(\ell'+1)\Gamma(\ell+1)}}{\Gamma(J-\ell+1)\Gamma(J-\ell'+1)\Gamma(J-\ell''+1)}. \end{aligned} \tag{37}$$

It is obvious that in the limit $n \to \infty$ the leading term corresponds to $\ell'' = 0$. Due to the triangular relation the condition $\ell'' = 0$ also implies $\ell' = \ell$. Therefore, in essence, we end up to leading order in $1/n$ with the free evolution:

$$a_\ell(j+1) \approx \exp\left\{-\frac{i\hbar\tau}{2I}\,\ell(\ell+n-1)\right\} a_\ell(j). \tag{38}$$

This fact may be explained as follows. Whereas in the classical limit the motion takes places in a two-dimensional plane, say the $(x_n, x_{n+1})$-plane, quantum mechanics allows also for fluctuations in the other degrees of freedom $(x_1, \ldots, x_{n-1})$. Hence, in the large $n$ limit the quantum model may be viewed as a planar kicked quantum rotator coupled to $(n-1)$ harmonic oscillators. The increase, respectively, the decrease in the energy is compensated by these oscillators.

### B. Numerical results for $\mathcal{M} = \mathcal{O}$ and finite $n$

We have numerically iterated the map (35) for various parameter sets and various initial states $|\Psi_0\rangle$. From the resulting angular-momentum distribution $|a_\ell(j)|^2$ we have calculated the expectation value of the kinetic energy,

$$\begin{aligned} E(j) :=&\ \langle\Psi_j| - (\hbar^2/2I)\Delta|\Psi_j\rangle \\ &\approx \frac{\hbar^2}{2I}\sum_{\ell=0}^{\Lambda}\ell(\ell+n-1)|a_\ell(j)|^2, \end{aligned} \tag{39}$$

and the coarse-grained angular-momentum distribution

$$A_\ell(j_{max}, \Delta j) := \frac{1}{\Delta j}\sum_{j=j_{max}+1}^{j_{max}+\Delta j}|a_\ell(j)|^2. \tag{40}$$

In the above $\Lambda$ is a cutoff parameter and $j_{max}$ and $\Delta j$ are chosen such that (40) results in a quasi-stationary coarse-grained angular-momentum distribution, that is, it becomes independent of $j_{max}$. As the long-time behavior of the above quantities is, in essence, independent of the initial state we present only results for $|\Psi_0\rangle = |0\mathcal{O}\rangle$.

As in the one-dimensional case we observe resonances, that is

$$E(j) \propto j^2, \tag{41}$$

if the quantity $(\hbar\tau/2I)$ is a rational multiple of $2\pi$. A typical resonance is shown in Figure 6 where the units have been chosen to $I = \hbar = 1$ and $\tau = 4\pi$. Here, the angular-momentum distribution $|a_\ell(j)|^2$ does not become quasi-stationary. With increasing time higher and higher angular-momentum eigenstates are excited. After, say, 500 kicks a cutoff parameter $\Lambda = 7000$ is no longer sufficient. The numerical iteration is no longer practicable.

For the non-resonance case we used units $\hbar = \tau = I = 1$, whence $K = \lambda = \kappa$. A small kicking strength, that is, $K < 4.5$, has been found to lead always to quasi-periodic oscillations in the energy (39). See Figure 7 for a typical example. Only the lowest angular-momentum eigenstates are excited and, hence, an angular-momentum cutoff $\Lambda \approx 20$ is sufficient.

For a larger kicking strength the energy (39) initially increases linearly in time, displaying the classical behavior (11). After about 20-50 kicks the quantum expectation value deviates from the classical one and shows fluctuations about a constant value (see Figure 8). This phenomenon is called dynamical localization and is known from the one-dimensional quantum rotator. Indeed, the corresponding state $|\Psi_j\rangle$ is localized in (angular) momentum space. Its coarse-grained angular-momentum distribution (40) becomes quasi-stationary, for sufficiently large $j_{max}$ and shows an exponential decay in $\ell$:

$$A_\ell \sim \exp\{\alpha\ell\}. \tag{42}$$

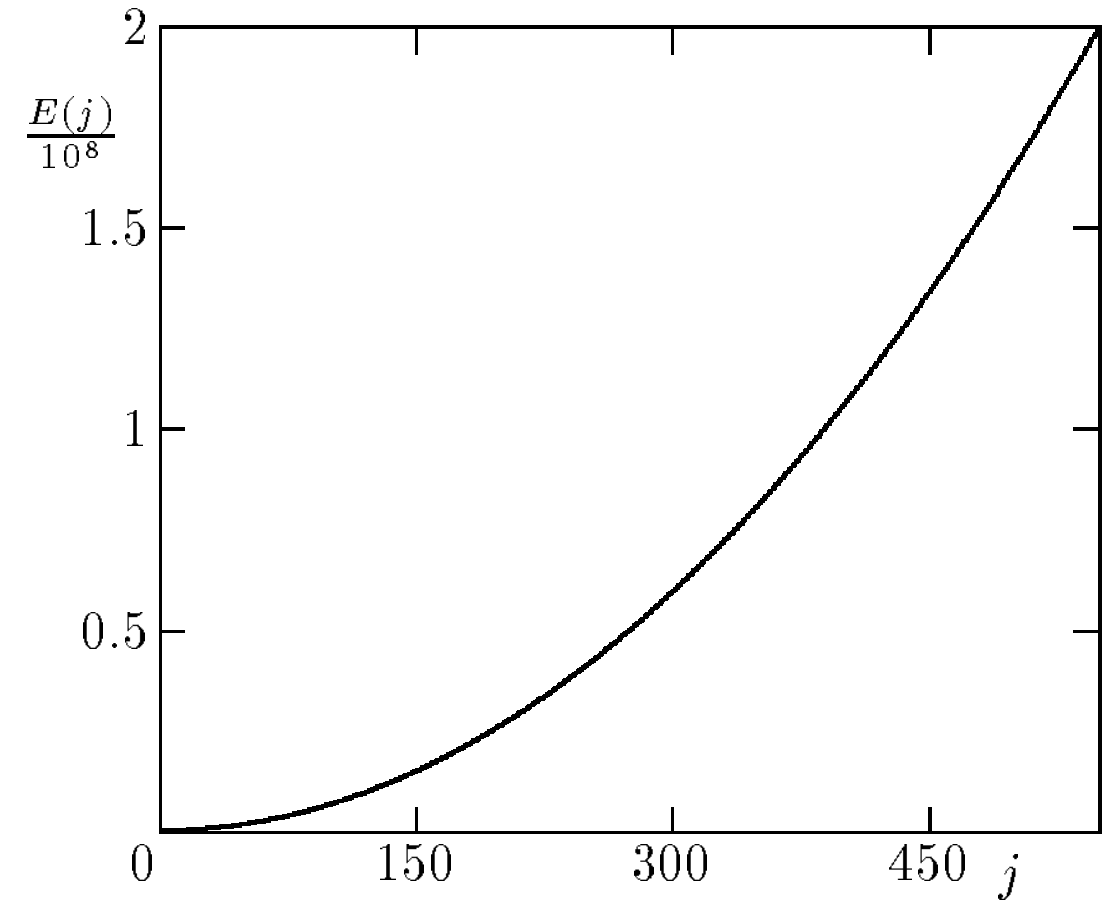

FIG. 6. Expectation value (39) of the kinetic energy for a resonance. Parameters are $\hbar = I = \lambda = 1$, $\tau = 4\pi$ and $n = 2$.

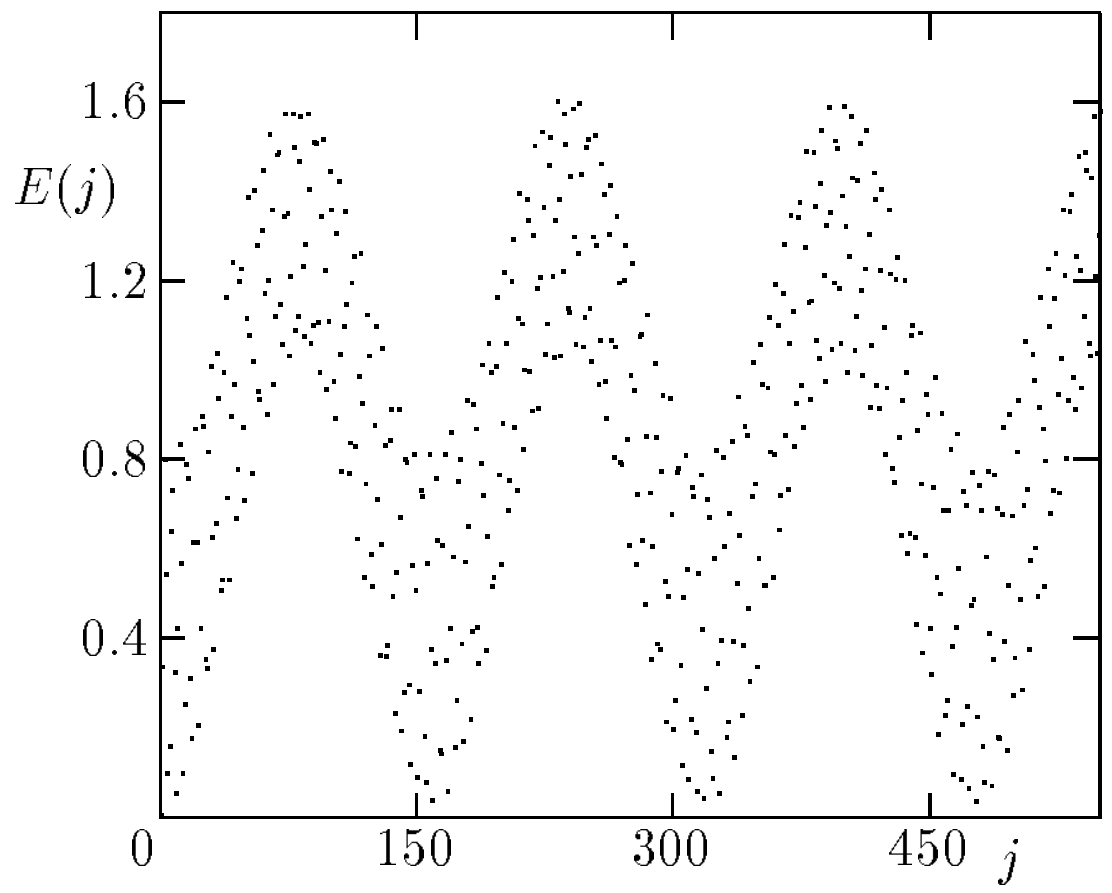

FIG. 7. Same as Figure 6 for the quasi-periodic case $\hbar = \tau = I = \lambda = 1$ and $n = 2$.

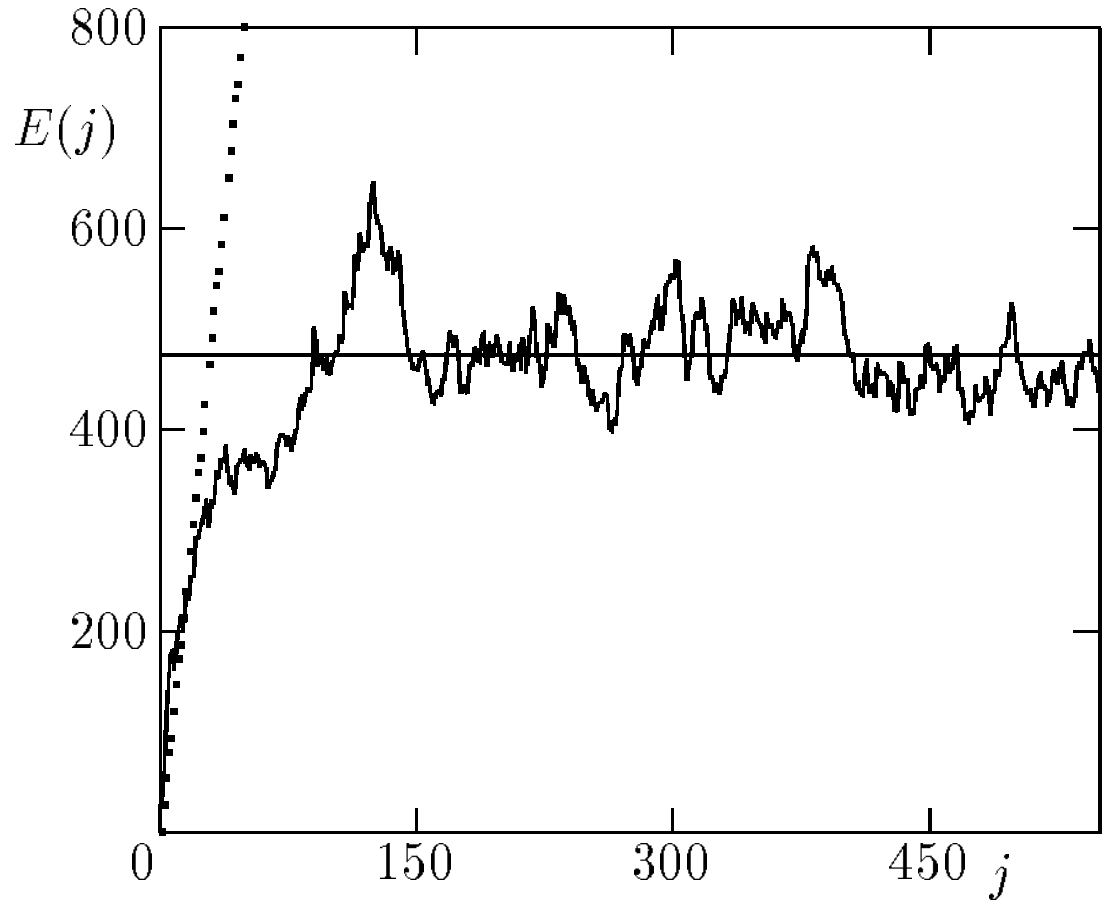

FIG. 8. Same as Figure 6 for parameters $\hbar = \tau = I = 1$, $\lambda = 9$ and $n = 2$. Here $\lambda$ is sufficiently large to lead to localization. The dotted line indicates the classical diffusion according to (11).

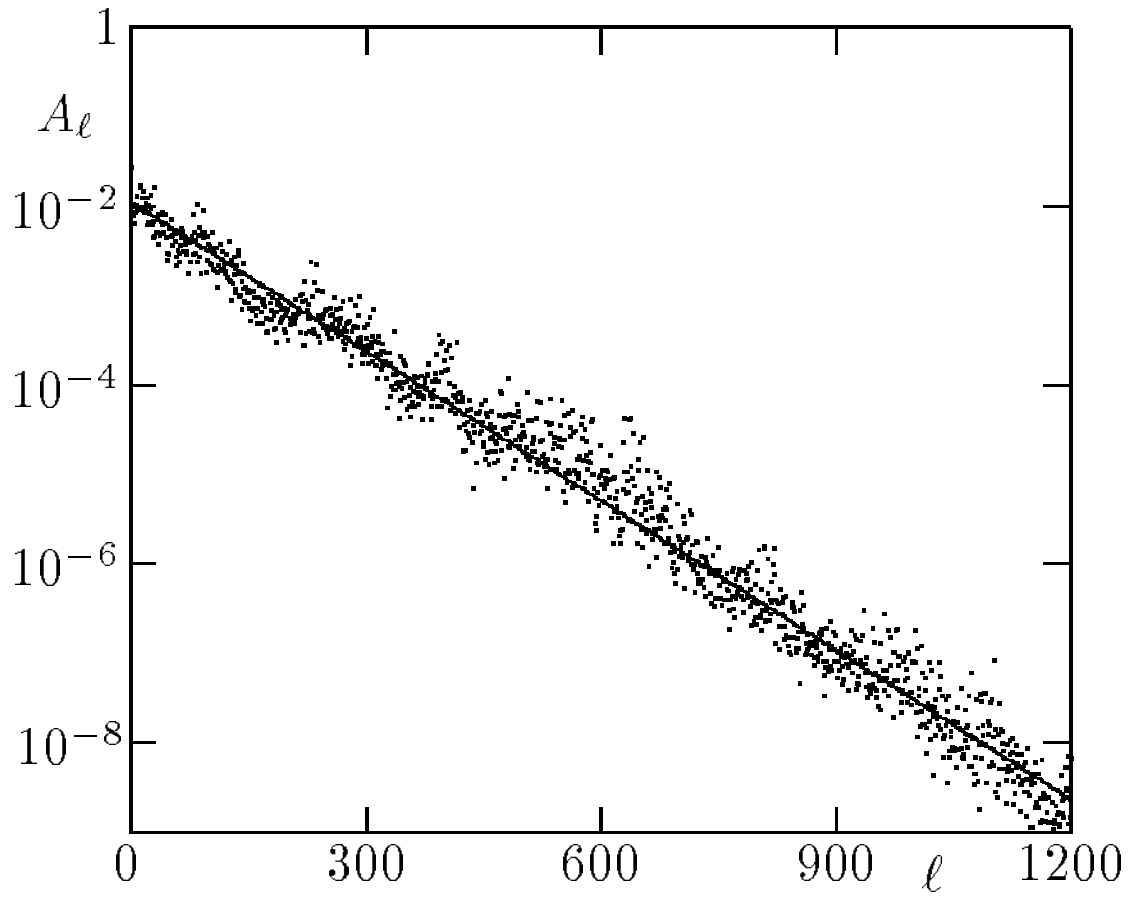

FIG. 9. The coarse-grained angular momentum distribution $A_\ell$ for parameters $\hbar = \tau = I = 1$, $\lambda = 14$ and $n = 2$. The localization is indicated by the solid line which is a fit to the exponential decay (42).

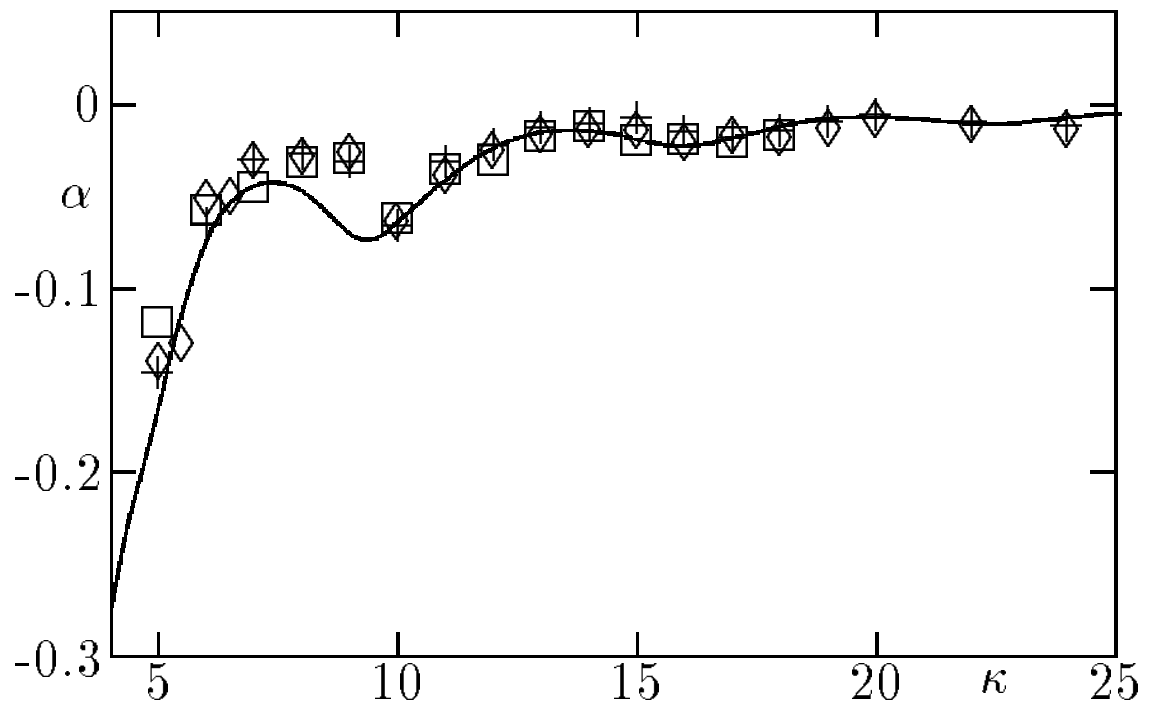

FIG. 10. The negative inverse localization length $\alpha$ as a function of $\lambda = \kappa$ (units $\hbar = \tau = I = 1$) obtained from a fit of (42) to our numerical data for $n = 2$ ($\Diamond$), $n = 4$ (+) and $n = 16$ ($\Box$). The solid line displays the analytical behavior (43).

The constant $\alpha$ is the negative inverse of the so-called localization length. A typical coarse-grained angular-momentum distribution is shown in Figure 9 for which we have chosen $j_{max} = 500$ and $\Delta j = 50$ in (40). The solid line is a fit to the exponential behavior (42). According to Shepelyansky [14] one expects the localization length to be identical with the classical diffusion constant (11), that is,

$$\alpha = -1/D(\kappa). \tag{43}$$

Despite the fact that Shepelyansky has derived (43) for the $n = 1$ case we find that this relation also holds for $n > 1$. In Figure 10 we compare the analytical behavior (43), where $D(\kappa)$ is given by (13), with our numerical results obtained from a fit to (42) for various values of $n$.

The $n$-independence of the localization length is explicitly visible in Table I where we show numerical results for $\alpha = \alpha(n)$ up to $n = 26$. Within this range of $n$ we do not see any signature of a transition to the free motion expected from the limit $n \to \infty$ as discussed in the previous Section.

## V. CONCLUDING REMARKS

In this paper we have studied the dynamics of an $n$-dimensional generalization of the kicked planar rotator. Classically, this system is found to have the same stochastic properties, that is diffusion in phase space, as those known from the well-studied case $n = 1$. This is not surprising as the classical dynamics does not depend on the number of degrees of freedom.

This is in contrast to the quantum dynamics which explicitly depends on this number $n$. Indeed, taking the naive limit $n \to \infty$ one recovers to leading order the free quantum motion [15]. That is, the kicks are lost in the sea of the infinitely many degrees of freedom. However, signatures of this disappearance have not been found by us for any finite value of $n$ which we have investigated. It turned out that we are limited by $n \leq 26$ for an explicit numerical calculation. In this range for the number of degrees of freedom we found no significant dependence on $n$. What we did find are resonances, quasi-periodicity and localization known from the quantum dynamics of the kicked planar rotator. It is still unclear, whether the transition from dynamical localization to free motion takes place at a finite value of $n$ or only in the limit $n \to \infty$ [16]. So more work is clearly needed for a fuller understanding.

## ACKNOWLEDGEMENTS

We would like to thank Fritz Haake for clarifying discussions. This work was partially supported by the Schwerpunktprogramm "Zeitabhängige Phänomene und Methoden in Quantensystemen in der Physik und Chemie" of the Deutsche Forschungsgemeinschaft.

[1] G. Casati, B.V. Chirikov, F.M. Izraelev and J. Ford, in *Stochastic Behavior in Classical and Quantum Hamiltonian Systems*, eds. G. Casati and J. Ford, Lecture Notes in Physics **93**, (Springer, Berlin 1979) p.334.

[2] M.V. Berry, N.L. Balazs, M. Tabor and A. Voros, Ann. Phys. (NY) **122**, 26 (1979).

[3] A.M. Ozorio de Almeida, *Hamiltonian Systems: Chaos and Quantization*, (Cambridge Univ. Press, Cambridge, 1988).

[4] M.C. Gutzwiller, *Chaos in Classical and Quantum Mechanics*, (Springer, New York, 1990).

[5] F. Haake, *Quantum Signatures of Chaos*, (Springer, Berlin, 1991).

[6] B.V. Chirikov, in *chaos and quantum physics*, eds. M.-J. Giannoni, A. Voros and J. Zinn-Justin, Les Houches 1989 Session LII, (North-Holland, Amsterdam, 1991) p.443.

[7] A.J. Lichtenberg and M.A. Liebermann, *Regular and stochastic motion*, (Springer, New York, 1983).

[8] F. Haake, M. Kuś and R. Scharf, Z. Phys. **B65**, 381 (1987). M. Kuś, R. Scharf and F. Haake, Z. Phys. **B66**, 129 (1987). R. Scharf, B. Dietz, M. Kuś and F. Haake, Europhys. Lett. **5**, 383 (1988).

[9] A.B. Rechester and R.B. White, Phys. Rev. Lett. **44**, 1586 (1980). A.B. Rechester, M.N. Rosenbluth and R.B. White, Phys. Rev. A **44**, 264 (1981).

[10] Y.H. Ichikawa, T. Kamimura, T. Hatori and S.Y. Kim, Prog. Theor. Phys. Supp. **98**, 1 (1989).

[11] N.J. Vilenkin, *Special functions and the theory of group representation*, (American Mathematical Soc., Providence, 1968).

[12] For details see A.O. Barut and R. Raczak, *Theory of group representations and applications*, (Polish Scientific Publ., Warszawa, 1980), second revised edition, p.290f.

[13] G. Junker, J. Phys. A **26**, 1649 (1993).

[14] D.L. Shepelyansky, Physica **28D**, 103 (1987).

[15] The discussion of Section IV.A indicates that the large $n$ limit corresponds to the weak-coupling limit $\lambda \to 0$.

[16] We may also note that the limit $n \to \infty$ becomes non-trivial if one simulaneously increases the kicking strength $\lambda$ such that the ratio $\lambda^2/n$ remains finite.

TABLE I. The negative inverse localization length as a function of the number of degrees of freedom $n$. Parameters are $I = \hbar = \tau = 1$ and $\kappa = \lambda = 11$. Although these numerical values are systematically above the theoretical value $\alpha = -4.1248 \times 10^{-2}$ expected from eq. (43), they do not indicate any $n$-dependence of $\alpha$.

| $n$ | 1 | 2 | 4 | 6 | 8 | 10 | 12 | 14 | 16 | 18 | 20 | 22 | 26 |
|---|---|---|---|---|---|---|---|---|---|---|---|---|---|
| $-\alpha \times 10^2$ | 2.5658 | 3.6280 | 3.0773 | 2.1610 | 2.8914 | 2.3077 | 3.1164 | 2.9679 | 3.2618 | 3.3897 | 3.0227 | 2.5675 | 2.9807 |